\documentclass[reprint,prl,superscriptaddress,amsmath,amssymb,aps,
]{revtex4-2}

\usepackage{bm}%
\usepackage{wasysym}
\usepackage{amssymb}
\IfFileExists{bbold.sty}{\usepackage{bbold}}{}
\usepackage[colorlinks=true,linkcolor=blue]{hyperref}%
\expandafter\ifx\csname package@font\endcsname\relax\else
 \expandafter\expandafter
 \expandafter\usepackage
 \expandafter\expandafter
 \expandafter{\csname package@font\endcsname}%
\fi
\usepackage{graphicx}
\usepackage{dcolumn}

\IfFileExists{ulem.sty}{\RequirePackage[normalem]{ulem}}{%
  }
\RequirePackage{color}

\newcommand{\safeincludegraphics}[2][]{%
  \IfFileExists{#2}{\includegraphics[#1]{#2}}{%
    \fbox{\parbox[c][4.2cm][c]{0.92\linewidth}{\centering
    Original figure file not supplied with the manuscript source\\[2mm]
    \texttt{\detokenize{#2}}}}}}

\begin{document}

\title{Parametric Glassiness in Graphene-based Electronic Lévy Glass}

\author{Caue N. da Silva}
\affiliation{Departamento de F\'{\i}sica, Universidade Federal Rural de Pernambuco, Recife - PE, 52171-900, Brazil}

\author{Diego B. Fonseca}
\affiliation{Departamento de F\'{\i}sica, Universidade Federal de Pernambuco, Recife - PE, 50670-901, Brazil}

\author{Adauto J. F. de Souza}
\affiliation{Departamento de F\'{\i}sica, Universidade Federal Rural de Pernambuco, Recife - PE, 52171-900, Brazil}

\author{Ant\^onio~M.~S.~Mac\^edo}
\email{antonio.smacedo@ufpe.br}
\affiliation{Laborat\'orio de F\'{\i}sica Te\'orica e Computacional,
Departamento de F\'{\i}sica, Universidade Federal de Pernambuco, Recife-PE 50670-901, Brazil}

\author{Luiz Felipe C. Pereira}
\email{luiz.cpereira@ufpe.br}
\affiliation{Departamento de F\'{\i}sica, Universidade Federal de Pernambuco, Recife - PE, 50670-901, Brazil}

\author{Anderson L. R. Barbosa}
\email{anderson.barbosa@ufrpe.br}
\affiliation{Departamento de F\'{\i}sica, Universidade Federal Rural de Pernambuco, Recife - PE, 52171-900, Brazil}

\begin{abstract}
Complex disordered conductors can display highly structured and
parameter-sensitive transport even in the absence of a thermodynamically stable
glass phase. Overlap distributions originally developed for replica symmetry
breaking (RSB) provide a useful diagnostic of this organization, provided
that their interpretation is distinguished from the equilibrium Parisi
construction. In this study, we report RSB-like overlap phenomenology and parametric mesoscopic
glassiness in a graphene-based electronic Lévy glass, a device that
mimics optical phenomena via electronic processes. By tuning the Fermi
energy, we can control the number of propagating wave modes and observe
reentrant changes from narrowly centered to broad or bimodal overlap distributions. The phenomenology is observed in both graphene-based and
conventional two-dimensional electron-gas electronic Lévy glasses,
indicating its generality. Sample-resolved overlap distributions show that the effect is
intrinsic to each disorder landscape, whereas the normalized conductance
variance shows that the growth of the absolute variance is mainly caused
by the increasing mean conductance. We therefore interpret the overlap as
a measure of correlation among hierarchically organized transport
pathways and the reentrant behavior as evidence of pinning,
parameter-induced reorganizations, and mesoscopic chaos in coherent
transport.
\end{abstract}


\maketitle



{\it Introduction} - Replica symmetry breaking (RSB), initially introduced
to explain the complex and rugged energy landscapes of spin glasses,
fundamentally involves frustration as a key feature that goes beyond
disorder~\cite{RevModPhys.95.030501}. RSB reveals a hierarchical
organization of metastable thermodynamic states and their Gibbs weights, thereby
underpinning the equilibrium behavior of systems with quenched disorder
and, owing to its broad conceptual scope, has become a fundamental tool
across diverse areas of pure and applied science and engineering
\cite{doi:10.1142/13341}. It is increasingly relevant for understanding
systems ranging from condensed matter~\cite{PhysRevB.104.184432} and
neural networks~\cite{NeuralNetworks} to ultracold atomic
systems~\cite{x19r-pzyb} and disordered photonic
media~\cite{Ghofraniha_2015,Gomes2016Glassy}.

In the past decade, RSB phase transitions have been demonstrated on
optical platforms, specifically random lasers, which include fiber
lasers, optofluidic systems, plasmonic nanostructures, and
glass-ceramics~\cite{Ghofraniha_2015,Gomes2016,Coronel2022,Qi2023Replica,Antenucci_2015,Tommasi2016Robustness,Gomes2016Glassy,Gomes2021Recent,Pierangeli2017}.
These systems exhibit strong intensity fluctuations, mode competition,
and nontrivial correlations, often accompanied by turbulent-like
dynamics and Lévy statistics~\cite{Coronel2022,Gomes2016}.
In random lasers and nonlinear optical media, repeated shots at
fixed external conditions provide stochastic replicas while nonlinear
mode coupling, gain competition, and noise supply the ingredients needed
for an effective statistical-mechanical description ~\cite{CalvaneseStrinati2026}.
In contrast,
the identification of strict RSB in coherent electronic
transport is more delicate~\cite{PhysRevLett.122.030401}.
For fixed Fermi energy, electrostatic potential, and disorder, the
Landauer problem is linear and deterministic and does not by itself
define a Gibbs measure, an effective temperature, or multiple pure
states. Nevertheless, strong Anderson localization can possess an intrinsic
glassy phenomenology: in the directed-polymer regime, where disorder strength is an order of magnitude greater than the hopping parameter, transport paths
are pinned by disorder, reorganize through avalanches as a parameter is
varied, and exhibit chaos under weak
perturbations~\cite{PhysRevLett.122.030401,PhysRevB.72.024206,PhysRevLett.99.116602}.

In this context, graphene acts as an exceptional electronic platform for mimicking optical phenomena through electronic processes~\cite{PhysRevB.87.155409,https://doi.org/10.1002/pssb.201552119,Caridad2016,PhysRevE.99.032118,PhysRevE.106.054127,PhysRevB.107.155432, PhysRevB.110.075421}. This is primarily due to its linear dispersion relation at low excitation energies, which qualitatively agrees with that of photons~\cite{Caridad2016}. 
The linear dispersion results from graphene's honeycomb lattice structure, which resembles a triangular Bravais lattice with a two-atom basis. Thus, charge carriers behave as massless relativistic Dirac fermions~\cite{RevModPhys.81.109}. 

\begin{figure*}
    \centering
    \safeincludegraphics[width=1.0\linewidth]{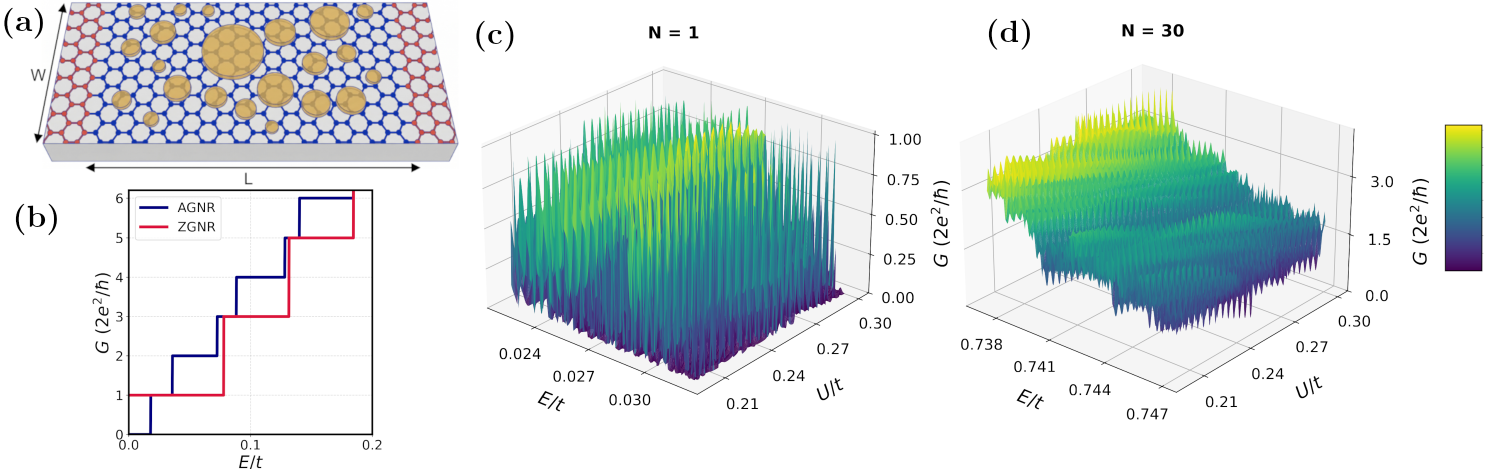}
    \caption{(a) Illustration of an ELG sample. It consists of an AGNR-based representation in blue and with length \( L \) and width \( W \), embedded in a random distribution of EQDs, illustrated in yellow. The sample is connected to two leads (shown in red) and deposited on a gray substrate. (b) The conductance \( G \) is plotted as a function of Fermi energy \( E \) (normalized by the hopping parameter \( t \)) for both pristine AGNR and ZGNR samples. In panels (c) and (d), the conductance energy spectrum through an AGNR-based ELG sample is presented as a function of electrostatic potential \( U \) for \( N = 1 \) and \( N = 30 \), where \(N\) is the number of propagating modes in the leads, respectively.}
    \label{fig:painel}
\end{figure*}


By harnessing the unique electronic properties of graphene,
Refs.~\cite{PhysRevB.107.155432, PhysRevB.110.075421} introduced the
concept of an electronic Lévy glass (ELG), an electronic analog of the
optical Lévy glass~\cite{Barthelemy2008}. In this system, charge
transport is governed by a scale-free distribution of electrostatic
quantum dots (EQDs), whose radii follow a power-law distribution, as
illustrated in Fig.~\ref{fig:painel}(a). As a consequence, electronic
trajectories span a broad range of length scales, leading to a transition
from superdiffusive transport at low Fermi energies, where only a few
propagating modes contribute, to diffusive transport at higher energies
with an increasing number of propagating modes. This behavior contrasts
sharply with conventional Anderson-disordered systems, where transport
is governed by spatially homogeneous, short-range
disorder~\cite{RevModPhys.80.1355}. In the ELG, the heavy-tailed geometry
of the EQDs creates a hierarchy of transport pathways
and tunneling channels, including bottlenecks,
resonant regions, and competing coherent routes. This hierarchy need not
constitute spin-glass frustration; instead, it defines a mesoscopic
transport landscape in which the dominant family of paths can be pinned
over a parameter interval and reorganized when a new channel or resonance
becomes available.

In this work, we use an RSB-inspired overlap to characterize parametric
glassiness in the conductance-energy spectrum (i.e., conductance versus
Fermi energy) through the ELG sample. More specifically, we show that at
low energy the spectra obtained at nearby
electrostatic potentials are weakly correlated and yield an overlap
distribution centered near zero. However, increasing the energy
opens additional propagating modes and
produces broad or bimodal overlap distributions, with reentrant returns
to a distribution centered near zero for particular mode numbers. This
nonmonotonicity motivates a transport-landscape interpretation rather
than a thermodynamic temperature analogy. Beyond that, the electrostatic
potential applied to lattice sites by EQDs is weak. 

{\it Microscopic model} - We explore parametric
glassiness by analyzing the conductance through a graphene-based ELG
sample, which is designed as a graphene nanoribbon connected to two leads
and embedded in randomly located EQDs, Fig.~\ref{fig:painel}(a). 
We consider two types of graphene nanoribbons: armchair (AGNR) and zigzag (ZGNR) with length $L \approx 1050a$ and width $W \approx 50a$, where \(a= 2.49\)~\AA~is the graphene lattice constant.
The scattering matrix
that describes the electronic transport through the nanoribbons is given
by
\[
\bf{S} = 
\begin{bmatrix}
{\bf{r}} & {\bf{t}'} \\
{\bf{t}} & {\bf{r}'}
\end{bmatrix},
\]
where \(\bf{t} \) and \( \bf{t}' \) represent the transmission blocks, while \( \bf{r} \) and \( \bf{r}' \) denote the reflection blocks. The conductance can be calculated using the Landauer-Büttiker relation
\(
G = \left(2e^2/h\right)\text{Tr}[\bf{t} {\bf{t}}^\dagger], \label{T}
\)
which is valid in the linear-response regime and at low temperatures~\cite{Datta_1995}. We performed numerical calculations of the conductance using the KWANT software \cite{Groth_2014}, which employs Green's-function methods within the tight-binding framework.
The tight-binding Hamiltonian for graphene is defined as follows
\[
\hat{H} = -t \sum_{\langle i, j \rangle} c_{i}^\dagger c_{j} + \sum_{i} \epsilon_i c_{i}^\dagger c_{i}, \label{H}
\]
where the indices \( i \) and \( j \) span all lattice sites, and \( \langle i, j \rangle \) indicates first nearest neighbors. The first term in \( \hat{H} \) represents the typical electron hopping between lattice sites, with \( c_{i} \) and \( c_{i}^{\dagger} \) being the annihilation and creation operators, respectively. The hopping energy \( t \) typically has a value of 2.7~eV~\cite{RevModPhys.81.109}. 
The second term in \( \hat{H} \) accounts for the electrostatic potential induced by the EQDs. Accordingly, the on-site electrostatic potential \( \epsilon_i \) equals \( U \) when the site is within the area of the EQDs, and vanish otherwise.
Experimentally, \(U\) can be controlled by applying a gate voltage to the substrate \cite{Caridad2016}. 

Figure~\ref{fig:painel}(b) illustrates the conductance as a function of Fermi energy for AGNR and ZGNR at null electrostatic potential (\(U=0\)). As expected, the conductance is quantized in terms of energy, meaning conductance can only take on integer values $N$. This integer \(N\) represents the number of propagating wave modes in the leads and is proportional to the lead width (\(W\)) and the Fermi wave vector (\(k_F\)) according to the equation \(N = {k_F W}/{\pi}\). Each \(N\) value has a different energy range, or energy spectrum. For AGNR, \(N\) can assume both even and odd values \(N=1, 2, 3, 4, \dots\), while ZGNR only allows odd values \(N=1, 3, 5, 7, \dots\)~\cite{RevModPhys.81.109}.

The ELG is designed to exhibit anomalous transport properties arising from a specific disorder geometry. 
In this setup, EQDs centers are randomly distributed within the nanoribbon's dimensions, satisfying strict non-overlap constraints. 
The radii of these circles, denoted as \(R\), follow a power-law distribution given by \(P(R) \propto R^{-(\beta + 1)}\), with \(\beta=1.22\). 
This heavy-tailed radii distribution enables transition between superdiffusive and diffusive transport regimes~\cite{PhysRevB.107.155432, PhysRevB.110.075421}. For more details, see Supplement Material (SM)~\cite{SM}.

\begin{figure}
    \centering
    \safeincludegraphics[width=0.9\linewidth]{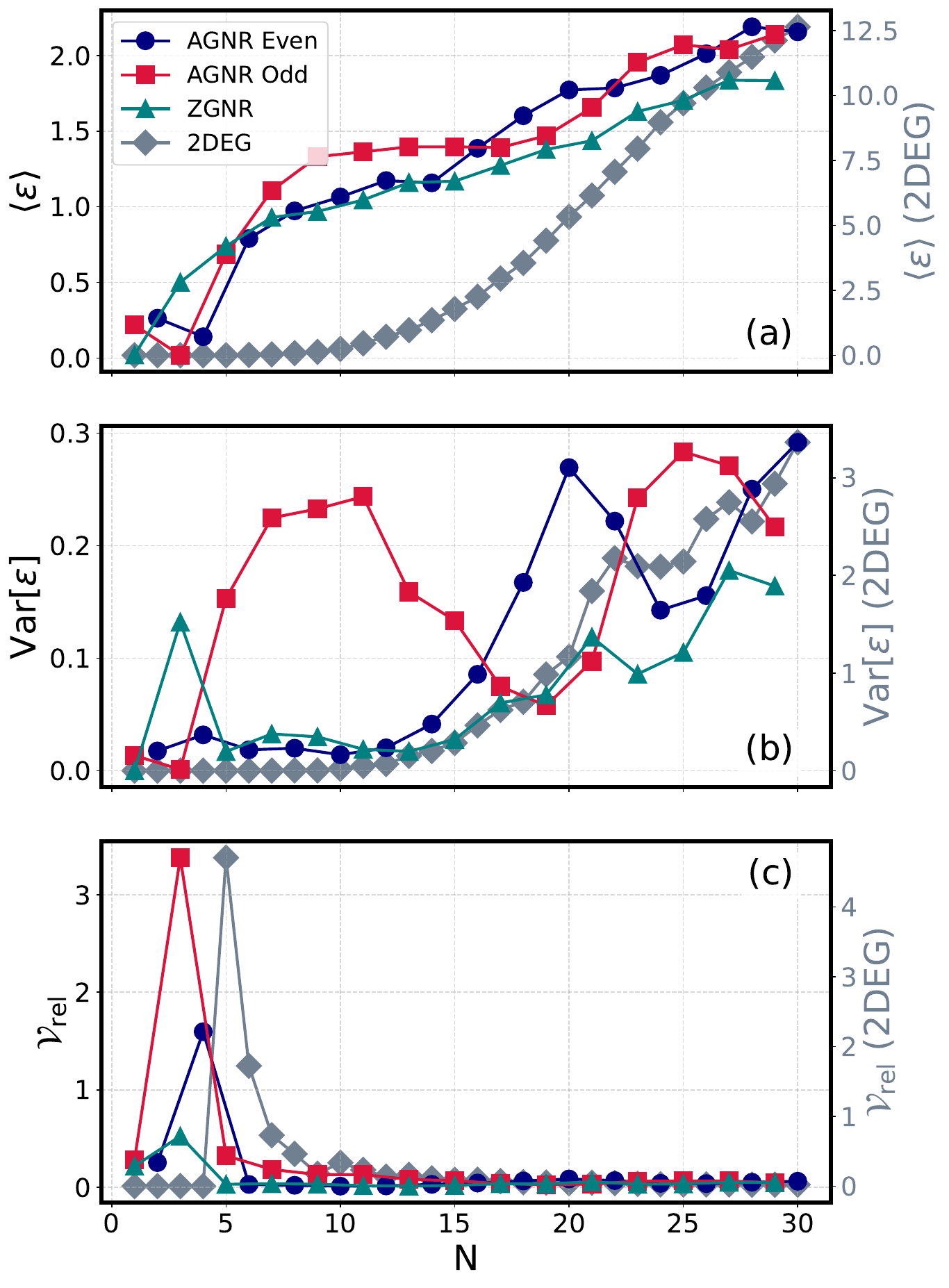}
    \caption{(a) Mean and (b) variance of the conductance energy spectral intensity, Eqs.~\ref{mg} and \ref{vg}, and (c) the dimensionless relative variance \( \mathcal{V}_{\mathrm{rel}} = {\operatorname{var}[\epsilon]}/{\langle\epsilon\rangle^2}\) versus propagating wave modes \(N\), for AGNR (even and odd modes separated), ZGNR, and 2DEG-based ELG samples. The left vertical axis uses the common scale for AGNR and ZGNR samples, while the right vertical axis uses the 2DEG scale.} 
   \label{fig:mean_var}
\end{figure}

Figures~\ref{fig:painel}(c) and ~\ref{fig:painel}(d) display the conductance energy spectrum through an AGNR-based ELG sample, varying the electrostatic potential  \(U=0.2t,\cdots,0.3t\) for \(N=1\) and \(N=30\), respectively. For \(N=1\), the energy spectrum shows random resonance peaks. In contrast, for \(N=30\), it is softer and no longer random. It is important to note that the range of the energy spectrum depends on the number of wave modes \(N\), as shown in Fig.~\ref{fig:painel}(b).
Furthermore, an ELG differs fundamentally from conventional Anderson disorder because transport develops through a hierarchy of competing pathways generated by the scale-free distribution of EQDs, as illustrated by the local density of states in SM~\cite{SM}.
The energy spectral resolution on the plateaus is \(\delta E = 10^{-3}t\) for any value of \(N\). 

Since the ELG sample disorder remains unchanged --- meaning the positions
and sizes of the EQDs do not change --- each energy spectrum measured
from a small variation of electrostatic potential
\(U\) is referred to below as a {\it parametric replica}. This terminology denotes a family of
responses within the same quenched disorder landscape, but it must not be
identified with a thermodynamic replica: changing \(U\) changes the
Hamiltonian, and the spectrum is deterministic when \(E\), \(U\), and the
EQD configuration are fixed. For a single sample, we numerically compute
101 energy-spectrum parametric replicas, each with 300
energy steps for each wave mode \(N\), as shown in
Figs.~\ref{fig:painel}(c,d). To obtain more reliable results, we developed
three data sets from three different ELG samples for each \(N\) value.
This is essential to demonstrate that the overlap phenomenology is independent of the particular
EQD realization~\cite{SM}. 

\begin{figure*} 
    \centering
    \safeincludegraphics[width=0.9\linewidth]{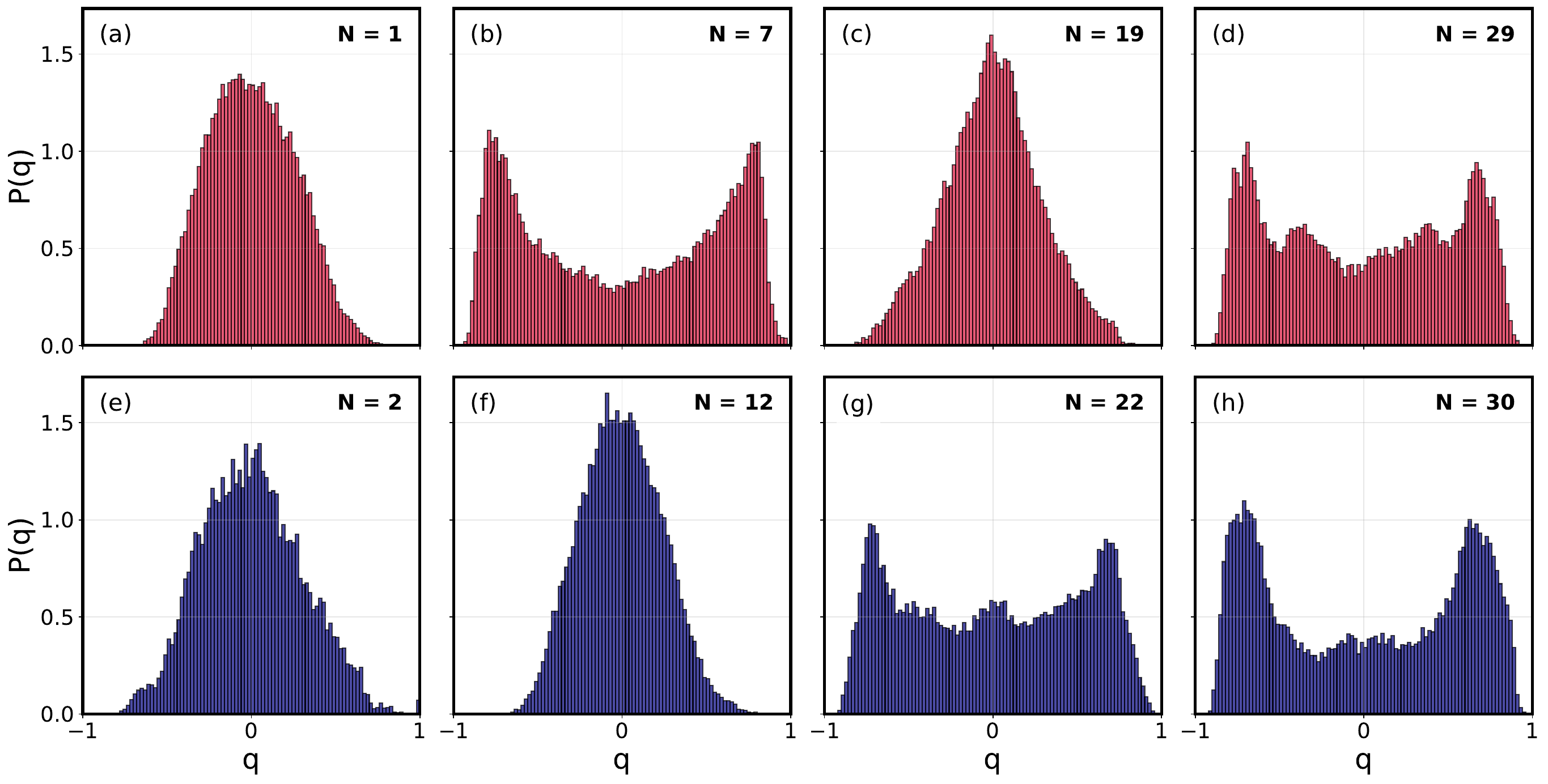}
    \caption{Histograms of the parametric-overlap distribution $P(q)$ of AGNR-based ELG samples for different wave modes $N$ ((a-d) odd modes in blue, (e-h) even modes in red), showing reentrant changes in spectral correlations as $N$ varies.}
   \label{fig:histogramas}
\end{figure*}

{\it Results} - We analyze the conductance energy spectrum to characterize correlations under parametric
changes of \(U\). 
{In our analysis, each energy spectrum corresponds to a distinct Hamiltonian response for the same quenched EQD configuration, arising from a near-parameter change.}
The average is calculated over an equal number
of wave modes, \(N\), with the same Fermi energy range.
The average conductance energy spectrum is defined as
\(
\epsilon^k_\alpha = \frac{1}{M} \sum_{i=1}^M g^k_\alpha(i),
\)
where \(g^k_\alpha = G^k_\alpha/\left(2e^2/h\right)\) represents the dimensionless conductance, \(\alpha\) is the parametric replica index, \(M = 300\) is the number of steps in the energy spectrum, and \(k\) is the sample index~\cite{Ghofraniha_2015}. The variation of this parameter, which depends on the electrostatic potential \(U\), quantifies the observed fluctuations. 
The mean and variance of \(\epsilon^k_\alpha\) over \(M_s = 101\) parametric replicas are given by
\begin{equation}
\langle \epsilon \rangle = \bigg\langle \frac{1}{M_s} \sum_{\alpha=1}^{M_s} g^k_\alpha \bigg\rangle_{k},\label{mg}
\end{equation}
and 
\begin{equation}
    \text{var}\left[\epsilon\right] = \bigg\langle\frac{1}{M_s} \sum_{\alpha=1}^{M_s} \left( \epsilon^k_\alpha - \langle \epsilon^k \rangle\right)^2\bigg\rangle_{k},\label{vg}
\end{equation}
where \(\langle~~\rangle_k\) is the average over \(k=3\) different ELG samples.

Figure~\ref{fig:mean_var}(a) illustrates \(\langle \epsilon \rangle\) as
a function of \(N\), calculated from Eq.~\ref{mg}. The mean energy
increases monotonically for both AGNR and ZGNR-based samples, showing no
signs of a regime change in the energy spectrum. In contrast, the
variance of the energy, calculated from Eq.~\ref{vg}, exhibits a
noticeable change in behavior as a function of \(N\), Fig.~\ref{fig:mean_var}(b). For all cases,
\(\text{var}\left[\epsilon\right]\) starts near zero for small values of
\(N\) (low energy) and begins to increase rapidly with the increase of \(N\)
(high energy), while the mean conductance grows as additional channels open.
In the case of AGNR-based ELG, we observe two distinct fluctuation patterns: one for even \(N\) and another for odd \(N\). 
Additionally, \(\text{var}\left[\epsilon\right]\) demonstrates non-monotonic behavior, alternating between increases and decreases depending on the specific value of \(N\).

To separate the scale of the fluctuations from the growth of the
mean signal, we also evaluate the dimensionless relative variance
\(
    \mathcal{V}_{\mathrm{rel}} =
   {\operatorname{var}[\epsilon]}/
    {\langle\epsilon\rangle^2}.
\)
As shown in Fig.~\ref{fig:mean_var}(c),
\(\mathcal{V}_{\mathrm{rel}}\) is largest near the first channel openings
and becomes small in the many-mode regime. The increase of the absolute
variance in Fig.~\ref{fig:mean_var}(b) is therefore dominated by the
growth of the mean conductance rather than by a new fluctuation scale.
Thus, the overlap distribution, which probes the organization rather than the
absolute amplitude of spectral fluctuations, is the more informative
diagnostic.


To analyze the system's behavior and
to characterize the organization of the ELG conductance spectra, we
introduce a normalized parametric-correlation diagnostic: the overlap
of intensity fluctuations across different parametric
replicas. The variable, which captures the behavior of individual wave
modes, is defined as the intensity fluctuation at a given frequency
\(
\Delta^k_\alpha (i) = g^k_\alpha (i) - \langle g^k (i) \rangle,
\)
where \(\langle g^k (i) \rangle\) represents the average intensity of each mode over the parametric replicas
\(
\langle g^k (i) \rangle = \frac{1}{M_s} \sum_{\alpha=1}^{M_s} g^k_\alpha(i).
\)
The overlap between energy conductance fluctuations is defined as
\[
q^k_{\alpha\beta} = \frac{\sum^M_{i=1} \Delta^k_\alpha (i) \Delta^k_\beta (i)}{\sqrt{\sum^M_{i=1} \left(\Delta^k_\alpha(i)\right)^2} \sqrt{\sum^M_{i=1} (\Delta^k_\beta (i))^2}}.
\]
From the numerically calculated spectra, we obtain the set of all
\(M_s(M_s-1)/2\) values of \(q^k_{\alpha\beta}\) for each input energy
and sample \(k\). From the three different samples, we have a data set
\(q=\{\{q^1\},\{q^2\},\{q^3\}\}\), which allows us to determine the
distribution \(P(q)\). We retain \(P(q)\) as a measure of the correlation between
conductance-fluctuation patterns obtained at nearby values of \(U\).
Although its mathematical form is the same as the intensity overlap used
in optical RSB experiments, here it quantifies parametric similarity
within a deterministic coherent-scattering problem and is not, by
itself, the Parisi overlap between thermodynamic pure states.

\begin{figure}
    \centering
    \safeincludegraphics[width=0.9\linewidth]{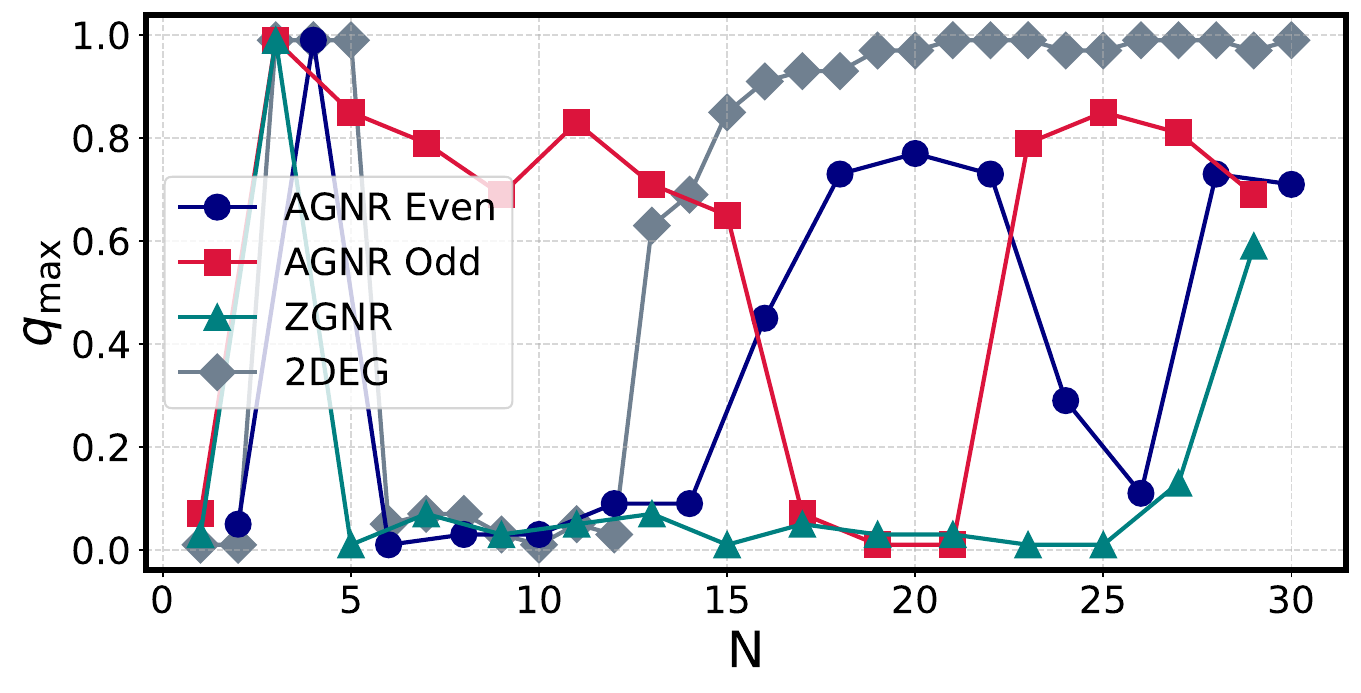}
    \caption{$q_{max}$ (values of $q$ which maximize the distribution $P( \lvert q  \rvert)$) versus wave modes $N$ for AGNR (even and odd modes), ZGNR, and the 2DEG-based ELG samples.}
    \label{fig:q_max}
\end{figure}

Figure~\ref{fig:histogramas} displays four examples of \(P(q)\) as the
number of wave modes \(N\) increases for both odd (a-d) and even (e-h)
modes in AGNR-based. For both cases, at low wave modes --- specifically
for \(N=1\) and \(N=2\) --- all overlap distributions are around zero,
indicating that the conductance-fluctuation patterns at nearby \(U\) values have no preferred positive or negative
correlation. As \(N\) increases, the overlap distribution \(q\) for odd
and even AGNR modes exhibits different behaviors, as shown in
Fig.~\ref{fig:mean_var}(b). 
For odd modes, as \(N \) increases to \(7\),
the overlap distribution becomes nontrivial, indicating the emergence of
strong correlations between parametric conductance spectra.
However, increasing to \(N=19\) results in the overlap distribution returning toward
zero, while the overlap distribution becomes nontrivial again at
\(N=29\), consistent with Fig.~\ref{fig:mean_var}(b). In contrast, for
even modes, the overlap remains centered near zero up to \(N=12\). Beyond \(N=12\), broad, structured correlations emerge.
All overlap distributions, \(N =1,\cdots,30\), can be  found in
SM~\cite{SM}. 
We also compute the Pearson correlation
coefficient~\cite{doi:10.1021/acsphotonics.7b01329,Coronel2022}, which is
small when \(P(q)\) is centered near zero and
larger when \(P(q)\) is broad or structured, consistent with overlap
distributions~\cite{SM}.


The sequence centered--broad--centered--broad displayed by the odd
AGNR modes is reentrant. Such behavior is difficult to reconcile with a
single thermodynamic transition driven monotonically by an effective
temperature, but it follows naturally from a mesoscopic landscape. Each
opening of a transverse subband changes the available modes,
bottlenecks, resonant regions, and coherent paths. For some \(N\), one
family of paths dominates over the sampled \(U\) interval and \(P(q)\)
is centered near zero; for others, nearby values of \(U\) redistribute
transport among correlated or anticorrelated families, producing a broad
or bimodal \(P(q)\). The heavy-tailed EQD distribution supplies a hierarchy
of length scales and resonances, making the selected family sensitive
and nonmonotonic in \(N\). In this sense, the observed behavior is
consistent with pinning, avalanche-like reorganizations, and mesoscopic
chaos in coherent electronic transport
\cite{PhysRevLett.122.030401}.

To support our findings, Fig.~\ref{fig:q_max} illustrates \(q_{max}\),
which corresponds to the position of the maximum of \(P(|q|)\) as a
function of \(N\) for AGNR and ZGNR. The value of \(q_{max}\) is
formally analogous to the Edwards-Anderson
order parameter in spin glass theory \cite{Antenucci_2015}.
Here, however, it locates the most probable parametric correlation
and is not assigned a thermodynamic order-parameter status. Unlike the
results reported by \cite{Ghofraniha_2015, Pierangeli2017,
Gomes2016Glassy}, where \(q_{max}\) remains nearly constant at the
saturation value of approximately \(1\), our results show that
\(q_{max}\) begins to decrease and increase at high \(N\). This behavior
is consistent with the findings presented in
Fig.~\ref{fig:histogramas} and aligns with the observations made
by~\cite{Gomes2016}. Its nonmonotonicity is another manifestation of
the reorganization of the active transport landscape as channels open.

To compare outcomes for graphene-based ELG samples, we performed
numerical calculations for an ELG sample in which the graphene lattice
is replaced by a square lattice, corresponding to a two-dimensional
electron gas (2DEG). Figure \ref{fig:mean_var}(b) shows that for the 2DEG,
the threshold in \(\text{var}\left[\epsilon\right]\) occurs at \(N = 10\),
which is similar to the behavior observed in both AGNR (even) and ZGNR.
Furthermore, the maximum value \(q_{max}\), as shown in
Fig.~\ref{fig:q_max}, indicates a change from weakly correlated spectra \(q_{max} \approx 0\) to strongly structured parametric correlations \(q_{max} \approx 1\)~\cite{SM}.
These results
demonstrate that the RSB-like overlap
phenomenology is not exclusive to graphene-based ELG systems; it can
also occur in different lattice structures. Therefore, the fundamental
cause of the observed organization is the scale-free
disorder geometry induced by random EQDs, rather than the Dirac
dispersion of graphene.

{\it Conclusions} - We have provided numerical evidence of parametric
mesoscopic glassiness in the electronic transport through a
graphene-based ELG. By systematically analyzing the conductance energy
spectra of AGNR and ZGNR embedded with randomly distributed EQDs, we
identified reentrant
changes between overlap distributions centered near zero and broad or
bimodal distributions as propagating channels are opened. The
sample-resolved analysis shows that the overlap
structure occurs within individual disorder realizations, whereas the
normalized variance demonstrates that the growth of the absolute
variance is mainly associated with the increase of the mean conductance.
Besides, the overlap phenomenology is reproducible across different EQDs configurations and is also observed
in a 2DEG-based ELG, confirming that it is driven by the scale-free hierarchy of paths, bottlenecks, and
resonant regions induced by the EQDs, rather than by the specific
Dirac-like dispersion of graphene. Because changing \(U\) changes
the Hamiltonian and no Gibbs ensemble or effective temperature is
defined, the present data should not be read as evidence for Parisi RSB
in the strict thermodynamic sense. Instead, \(P(q)\) quantifies whether
nearby parametric responses explore the same, anticorrelated, or
decorrelated families of coherent transport paths. The reentrance with
\(N\) is then a central result: it supports a landscape in which paths
are pinned over parameter intervals and reorganize through
avalanche-like events, with strong sensitivity to small perturbations.
ELG therefore provide a controllable platform for
studying intrinsic glassy phenomena of coherent localization and
scattering, including pinning, avalanches, and mesoscopic chaos.

\begin{acknowledgements}
CNS acknowledges a scholarship from Fundação de Amparo à Ciência e Tecnologia de Pernambuco (FACEPE, Grant IBPG-0981-1.04/25).
DBF acknowledges a scholarship from Coordenação de Aperfeiçoamento de Pessoal de Nível Superior (CAPES, Grant 0041/2022).
LFCP acknowledges financial support from CAPES (Grant 0041/2022), Conselho Nacional de Desenvolvimento Cient\'{\i}fico e Tecnol\'ogico (CNPq, Grants   200296/2023-0, 371610/2023-0 INCT Materials Informatics and 310262/2025-9).
ALRB acknowledges financial support from CNPq (Grants 302502/2025-4 and 406836/2022-1 INCT of Spintronics and Advanced Magnetic Nanostructures - SpinNanoMag).
AMSM acknowledges financial support from CNPq (Grants 310800/2026-9 and 307626/2022-9).
\end{acknowledgements}

\bibliography{ref.bib}


\end{document}